\documentstyle[12pt,aasms4]{article}

\font\MyScript=eusm10
\slugcomment{Astrophysical Journal}
\lefthead{Scargle}
\righthead{Bayes Blocks}
\renewcommand{\baselinestretch}{2}

\newcommand{\DS}{\renewcommand{\baselinestretch}{1.5} \tiny \normalsize}
\newcommand{\SS}{\renewcommand{\baselinestretch}{1}   \tiny \normalsize}

\begin{document}

\title{Studies in Astronomical Time Series Analysis:\\
V. Bayesian Blocks, A New Method to Analyze Structure in \\
Photon Counting Data }

\author{Jeffrey D.  Scargle}

\affil{Space Science Division\\
National Aeronautics and Space Administration\\
Ames Research Center}

\authoraddr{MS 245-3, NASA-Ames Research Center, Moffett Field, CA, 94035-1000}
%%\altaffiltext{1}{NASA/Ames Research Center, Moffett Field, CA, 94035-1000}

\authoremail{jeffrey@sunshine.arc.nasa.gov}


\SS

%========================================
\begin{abstract}
% A Bayesian analysis of photon-counting data
I describe a new time-domain algorithm for 
detecting localized structures (bursts),
revealing pulse shapes,
and generally characterizing intensity variations.
The input is raw counting data, in any of 
three forms:
time-tagged photon events (TTE), 
binned counts, 
or time-to-spill (TTS) data.
The output is the most probable 
segmentation of the observation 
into time intervals during which the photon arrival rate 
is perceptibly constant -- {\it i.e.} has no
statistically significant variations.
The idea is not that the source is deemed to 
have this discontinuous, piecewise constant form, 
rather that such an approximate and generic model is often useful.
Since the analysis is based on Bayesian statistics,
I call the resulting structures Bayesian Blocks.
Unlike most, this method does not stipulate time bins -- instead 
the data determine a piecewise constant representation.
Therefore the analysis procedure itself does not impose a lower limit 
to the time scale on which
variability can be detected.
Locations, amplitudes, and rise and decay times 
of pulses within a time series can be estimated, independent of any 
pulse-shape model -- but only if they do not overlap too much,
as deconvolution is not incorporated.
The Bayesian Blocks method is demonstrated by analyzing 
pulse structure in BATSE $\gamma$-ray data.
The MatLab scripts and sample data can be found on 
the World Wide Web at: 
\begin{verbatim}
http://george.arc.nasa.gov/~scargle/papers.html
\end{verbatim}
\end{abstract}

\vskip 1cm
\noindent
For information concerning US Government intellectual property
issues connected with the technology contained in this paper,
contact Jeanne Stevens, Commercial Technology Office,
NASA Ames Research Center, Mail Stop, 202A-3,
Moffett Field, CA  94035-1000, (650) 604-0065.

\tableofcontents
\DS

\newpage

%=========================================================
\section{The Problem: Structure in Photon Counting Data} 
\label{photon_counting} 
%---------------------------------------------------------

% the problem
Tracking a variable object's brightness 
changes, based on photon counting data,  
is a fundamental problem in astronomy.
For example, the importance of 
activity of galactic and extragalactic objects 
on time scales at and below the millisecond range  
led NASA to design its X-ray and $\gamma$-ray observatories to 
detect individual photons with microsecond timing accuracy.

%========================================
\subsection{Difficulties} 
\label{difficulties}
%----------------------------------------
% the difficulties
Existing methods do not fully and correctly 
extract the information in photon counts.
The scientifically useful information,
of course, is buried in the fluctuations inherent in 
the occurrence of discrete, independent events -- {\it i.e.}
photon detections.
The shortest time scales
are especially vulnerable to information losses.
There are at least three reasons for this.

First are the {\it binning fallacies}.
It is widely and incorrectly held that: 
(1) such data must be 
binned\footnote{{\it I.e.}, one must divide 
the observation into equally spaced intervals 
and count photons within these {\it bins}.}
in order to be analyzed at all, and  
(2) the bins must be 
large enough so that there are enough photons 
in each to provide a good statistical sample.
The almost universal practice of binning event data 
throws away a considerable amount of information 
and introduces dependency of the results on the
sizes and locations of the bins.

A second reason is that 
many analysts routinely use global methods, 
in essence averaging over the observation interval 
or subsegments of it that are sufficiently long
to provide a good statistical sample.
Power spectra, autocorrelation functions, 
and histograms are examples.
While good for some problems,  
global methods dilute short bursts 
or other local signals. 

Incorrect error models are the third source 
of information loss.
It is usually assumed 
that observational errors are additive and normally distributed
(as in $\chi^{2}$ methods).
Counting fluctuations are neither additive nor normal.
Indeed, {\it the nearly ideal Poisson nature of 
photon detection provides the rare advantage 
of knowing statistical properties of the noise 
with great confidence, completeness, and precision}.
(Typically the major way in which the data 
depart from this ideal is through lack of
independence.  In particular, 
detectors have a {\it dead-time} --
arrival of a photon momentarily
inhibits detection of subsequent photons.)

%========================================
\subsection{Approach} 
\label{approach}
%----------------------------------------

% the new approach
A single, simple idea sparked this development.
{\it The probabilities of the elementary events -- 
photon detection or nondetection --
have such a simple but exact specification
[equation (\ref{element}) below] that it ought 
to be easy to derive an explicit 
statistical treatment of the total problem.}
This led to a new algorithm, based on Bayesian principles, 
as described in \S \ref{analysis}
and demonstrated in \S \ref{example_batse}.
It exploits the full time resolution of the data, 
makes explicit use of the correct statistical distribution, 
avoids arbitrary binning, and operates in the time
domain -- focusing on local structures.  
It converts raw photon counts into 
the most probable piecewise constant representation of 
brightness as a function of time.
This decomposition can provide simple estimates of 
the width, location, and amplitude 
of pulses -- assuming their 
overlap is neglectable -- and of the background level,
without invoking parametric or 
other explicit pulse-shape models.
An excellent overview of Bayesian methods, with an astronomical 
flavor, is \cite{loredo}.
Readers unfamiliar with Bayesian time series analysis 
might consult \cite{sivia},  
or the overview, with 
specific discussion of the changepoint 
problem, in \cite{ruanaidh}.

% overview of goal
Before proceeding, a few comments on basic approach.
As is common in astronomy, 
the following conceptual scheme underlies the data analysis.
Some physical process in the astronomical object
causes brightness variations.
These fluctuations -- modified by radiative transfer,
viewing geometry, intervening matter, {\it etc.} --
are modeled as an idealized signal,
which in turn is compared with
one or more physical models of the original dynamical process.
Connection with the observed photon stream   
is made by interpreting the signal as determining  
a time-variable photon detection probability.
Mathematical properties of this function
({\it e.g.} smoothness or differentiability), 
correspond to physical properties of the
source -- some of which are known but others of
which are unknown.

In describing this kind of modeling,
terms like {\it pulse}, {\it burst}, and {\it shot}
have all been used, loosely, to mean more or less
the same thing -- namely a process that is in some
sense {\it local}, as opposed to {\it global}, in time.
I know of no generally accepted, rigorous definition
of any of these terms, but the following notions may 
be useful.
Consider a stochastic process with a continuous power spectrum 
of a simple functional form and 
extending over a broad range of time scales.
Call this the {\it global} process. 
Self-similar or ${1 \over f}$ processes are examples
[{\it cf}. \cite{scox,abry_2,young_scargle}].
A deterministic component with a line spectrum,
such as a periodic signal, may also be present without 
materially changing the picture.
{\it Bursts}, then, are non-periodic signals,  
localized in time, that are not part of the global process.
That is, the spectrum of the total signal 
is altered by the presence of the bursts and is not 
of the simple form postulated for the global signal.
Bursts can occur randomly, periodically, 
or in any other fashion.
In this picture, whether or not a statistical ensemble of 
signal features is deemed to be bursts depends on 
the events' shapes, distribution, and 
relation to the global signal.

% problems with global vs. local (theory/practice)
This distinction between global and local signals 
cannot always be made cleanly.
For example, {\it intermittency} in 
a chaotic nonlinear dynamical system 
[{\it e.g.} \cite{schuster}]
is in a sense localized, 
but is described by the same laws of motion 
which govern the chaotic behavior of the system.
Furthermore it is obvious that, in the presence of noise,
bursts can be detected only statistically.

The approach adopted here, 
using what statisticians call {\it change-point determination},
addresses part of this definitional problem head-on, 
as it is based directly on the statistical significance 
of putative local structure.
On the other hand, distributions of the times, amplitudes, or shapes
of pulses are not considered here;
these would be concerns of a follow-on study,   
after Bayesian block analysis of the full time series.

%========================================
\subsection{Other Work} 
\label{other_work}
%----------------------------------------

It has long been recognized that 
Bayesian methods are well-suited to 
finding changepoints
[\cite{smith,worsley}].
A Bayesian analysis of Poisson data 
similar in spirit to the present work
is \cite{raftery_akman};
see also Appendix C of \cite{gregory_loredo}.
\cite{west_ogden} use methods similar to those
described here 
to find changepoints in binned data,
to an accuracy better than the bin size.
(Their solutions are simultaneous maximum
likelihood in the rates and changepoint location;
the rate marginalization carried out here
is probably preferable.)
\cite{sugiura_ogden} discuss detection 
of gradual, linear trends, rather than sudden changes.

Localized basis functions, such as wavelets, 
provide a partial solution to this problem
[\cite{abry_1,wavelet_methods,brill_1}].
And the procedure described in \cite{eric}
is somewhat related to the present approach;
his segmentations are the standard 
dyadic intervals of wavelets,
whereas here the intervals 
adapt themselves to the data
and are therefore not generally evenly spaced.
\cite{donoho_1} studied
edge location in, and multi-segmented
analysis of, time series. 
His methods, {\it segmentation pursuit} 
and {\it minimum entropy segmentation}, 
circumvent the fixed location of
conventional wavelet methods,
for a more general 
statistical model that that used here.
Translation invariant 
wavelet transforms  
(\cite{coifman_donoho})
also have potential 
for accurate location of changepoints.

\cite{abry_2} discuss the
other side of the coin from the topic of the
current paper,
namely {\it long-range dependence} 
in point processes (the statistical term 
for event data, such as photon counting),
using wavelet methods.
Recent work has applied wavelets and wavelet denoising 
to the changepoint problem -- see 
\cite{ogden_1,ogden_parzen_a,ogden_parzen_b}.

I have recently become aware of the following
work, closely related to this problem: 
\cite{stark,gustafsson_1,gustafsson_2}.

%===============================================
\section{The Analysis Method: Bayesian Blocks}
\label{analysis}
%-----------------------------------------------

This section details  
a new algorithm implementing a Bayesian approach 
to the problem of detecting variability 
in photon counting data.
A sketch of standard Bayesian 
model fitting will set notation and the context.
We have some data $D$, and a model {\MyScript M} 
containing a parameter $\theta$.
If there are several parameters, simply interpret $\theta$
as a vector.
We want to estimate how probable it is 
that the model is correct, 
and we want to learn 
something about likely values 
of the parameter -- all based on the data and any 
prior information that we might have.

The basic relation quantifying parameter inference is 
Bayes' Theorem, one form of which is  
\begin{equation}
P( \theta | D, \mbox{\MyScript M} ) 
P( D | \mbox{\MyScript M} )
= P( D | \theta, \mbox{\MyScript M} ) 
P( \theta | \mbox{\MyScript M} ).
\label{bayes}
\end{equation}
\noindent 
In order, the conventional names of the factors are
the {\it posterior probability density} of $\theta$, 
given the data,
and the {\it prior predictive probability} for the data,
on the left side; and
the {\it likelihood} for the parameter,
and the {\it prior} probability of the parameter, 
on the right side.
These factors have other names to connote different emphasis; {\it e.g.} 
$P( D | \mbox{\MyScript M} )$
is sometimes called the {\it global or marginal likelihood for the model}. 
Also, as described by \cite{jaynes},
 $P( D | \theta, \mbox{\MyScript M} )$ is termed the likelihood 
when emphasizing its dependence on $\theta$, but
as the {\it sampling distribution} when emphasizing its dependence on $D$.
All of the terms are to be interpreted {\it given the model};
this is the meaning of {\MyScript M } behind the ``$|$''.
The two sides of this equation are simply different ways of
reckoning the probability of the same compound event,
{\it i.e.} the model parameter having a specific value 
and the data being as observed.
Standard practice is to write $P( D | \mbox{\MyScript M} )$
as a divisor on the right-hand side, as this is the 
way Bayes'  theorem is actually used: 
$P( \theta | D, \mbox{\MyScript M} )$ is the 
probability distribution of the parameter and
serves the role of quantifying the model's ``goodness of fit''
to the data.

%---------------------------------------------
\subsection{Comparison of Alternative Models}
\label{compare}
%---------------------------------------------

A key tool is 
a procedure to decide 
which of two (or more) alternative models
of a given chunk of data is more probable. 
This selection is based on those data 
plus any prior information  
on the relative likelihood of the models.
{\it E.g.}, we might want to choose between  
the following two models of an astronomical light curve, 
based on observations over a time interval $T$
\footnote{In what follows we use this symbol for both
a time interval and its length; this should not cause confusion. }:
\begin{itemize}
\item {\bf {\MyScript M}$_{1}$ }: constant intensity over $T$ 
\item {\bf{\MyScript M}$_{2}$ }: possibly different constant intensities
                     in two sub-intervals, $T_{1} + T_{2} = T$
\end{itemize}
\noindent
As will become apparent, 
this example is at the heart of the method proposed here.

Consider a set of $K$ models, say
{\MyScript M}$_{1}$, {\MyScript M}$_{2}$, {\MyScript M}$_{3}$, $\dots$, {\MyScript M}$_{K}$.
(By {\MyScript M}$_{k}$ we mean the model 
without specification of any parameter values,
so the terms model {\it class} or {\it structure} are better.)
That we are limiting consideration 
to this set, plus all other relevant knowledge or assumptions, 
together comprise a background of information, conventionally denoted $I$.
Bayes' theorem for model selection -- 
as opposed to parameter estimation as in equation (\ref{bayes}) --
gives for the posterior probability of each model,
given the data $D$ and the background information $I$, 
\begin{equation}
P( \mbox{\MyScript M}_{k} | D, I ) = 
{ P( D | \mbox{\MyScript M}_{k}, I ) P( \mbox{\MyScript M}_{k} | I ) 
\over P( D | I ) }
\label{bayes_evidence}
\end{equation}
\noindent
Since $I$ does not change, and is therefore irrelevant 
in comparisons of the kind considered here,
we often omit the symbol; its presence should be assumed in all 
equations derived from Bayes' theorem, including
equation (\ref{bayes}).

Equation (\ref{bayes_evidence}) immediately gives 
a comparison of how well two models 
represent the data, in terms of the {\it odds ratio} 
\begin{equation}
{P( \mbox{\MyScript M}_{k} | D ) \over P( \mbox{\MyScript M}_{j} | D) } = 
{ P( D | \mbox{\MyScript M}_{k} ) P( \mbox{\MyScript M}_{k} )
\over
P( D | \mbox{\MyScript M}_{j} ) P( \mbox{\MyScript M}_{j} ) }
\label{compare_init}
\end{equation}
Note that $P( D | I )$ -- the probability of observing the data, 
without regard to the model -- is irrelevant to comparison of model classes
and accordingly cancels out.

The quantity 
$P( D | \mbox{\MyScript M}_{k} )$, 
the probability of the data given the model,
can be found by integrating equation (\ref{bayes}) 
over $\theta$, making use of the fact that 
$P( \theta_{k} | D, \mbox{\MyScript M}_{k} )$
is normalized:
\begin{equation}
P( D | \mbox{\MyScript M}_{k} ) = 
\int P( D | \theta_{k}, \mbox{\MyScript M}_{k} ) 
P( \theta_{k} | \mbox{\MyScript M}_{k} ) d\theta_{k}, 
\label{global_like}
\end{equation}
\noindent
The number and significance of the parameters may be different 
from model to model -- hence the subscript on $\theta_{k}$.
Thus equation (\ref{compare_init}) becomes
\begin{equation}
{ P( \mbox{\MyScript M}_{k} | D ) 
\over 
P( \mbox{\MyScript M}_{j} | D) } = 
{ \int P( D | \theta_{k},  \mbox{\MyScript M}_{k} ) 
P( \theta_{k} | \mbox{\MyScript M}_{k} ) d\theta_{k}
\over
\int P( D | \theta_{j},  \mbox{\MyScript M}_{j} ) 
P( \theta_{j} | \mbox{\MyScript M}_{j} ) d\theta_{j} } \ \
{ P( \mbox{\MyScript M}_{k} ) \over P( \mbox{\MyScript M}_{j} )}
\label{basic_with_priors}
\end{equation}
\noindent
From this equation it is clear that 
\begin{equation}
J( \mbox{\MyScript M}_{k}, D )
\equiv 
P( \mbox{\MyScript M}_{k} )
\int P( D | \theta_{k}, \mbox{\MyScript M}_{k} ) P( \theta_{k} | \mbox{\MyScript M}_{k} ) d\theta_{k}
\label{total_like}
\end{equation}
\noindent
is the fundamental quantity to be used in comparing models
($J$ for joint probability for the model and the data). 
% I therefore call this the {\it total likelihood} of the model:
This factor includes prior information, and 
is independent of the number of, or values of, any model parameters.
The model with the largest $J$ value
is the most likely to be correct.
The integral on the right side of
equations (\ref{global_like}) and (\ref{total_like}),
\begin{equation}
\mbox{\MyScript L}( \mbox{\MyScript M}_{k}, D ) = 
\int P( D | \theta_{k}, \mbox{\MyScript M}_{k} ) P( \theta_{k} | \mbox{\MyScript M}_{k} ) d\theta_{k}
\label{global_like_1}
\end{equation}
\noindent 
is often called 
the {\it global likelihood},
or sometimes the {\it marginal likelihood}
or the {\it evidence} for the model.

It is the essence of the problems considered here 
that we are ignorant 
about the different model structures prior to 
analyzing the data.
Accordingly the model priors $P( \mbox{\MyScript M}_{k} )$ 
(not to be confused with priors for the parameter) 
could all be taken equal and omitted from 
expressions for the global likelihood.
However, for practical reasons 
it is useful to retain the
{\it prior odds ratio}
\begin{equation}
\rho = { P( \mbox{\MyScript M}_{k} ) \over P( \mbox{\MyScript M}_{j} )}
\label{prior_odds}
\end{equation}
\noindent
as a scalar parameter of the computations.
In the sample applications described 
in \S \ref{example_batse} below
this quantity is used to suppress 
spurious blocks due to the
statistical fluctuations.

Note that the complexities of the models, 
{\it e.g.} the number of parameters,
are automatically accounted for in this comparison.
Adding parameters to a model almost always increases its maximum likelihood
(rigorously, never decreases it).
But as is well known, the best model is not the most complex one.
Some modeling techniques introduce a penalty factor 
that compensates for the added degrees of 
freedom represented by a more complex model.
Here, as usual in Bayesian analysis, 
{\it this tradeoff between goodness of fit and model
complexity is an automatic consequence of the 
integration over all model parameters} in equation (\ref{global_like_1}).
Sometimes in Bayesian analyses such a penalty factor 
is isolated and called the {\it Ockham factor}.
\cite{jaynes} has a nice discussion of this issue;
see also Chapter 4 of \cite{sivia}.

%---------------------------------------------------
\subsection{Evidence for a Constant Poisson Rate Model}
\label{evidence_single}
%---------------------------------------------------

Now let's use equation (\ref{global_like_1})
to compute the global likelihood  
for a source being of constant intensity
during a given observation interval. 
The {\it Poisson process}
is the mathematical model of such a source,
with $\lambda \ge 0$ denoting the rate, 
here in photons per unit time, 
assumed to be constant over some 
time interval $T$.
That is, the photon events are 
distributed identically,
independently of each other,
and with uniform probability over $T$
at rate $\lambda$ per unit time.
Think of drawing a random integer from the 
Poisson counting distribution with mean $\lambda T$
and then throwing this number 
of darts randomly and uniformly across the interval.
It is well known that this 
process has no memory or after effect
in waiting-times: the arrival of a photon
does not affect the probability of subsequent
photon arrivals. This property implies that 
waiting times have an exponential distribution
[\cite{billingsley}, \S 23].
The Poisson model therefore has zero dead-time.

We actually use the discrete-time 
version, the {\it Poisson counting process} (PCP).  
That is, the observation interval 
is divided into a number of equal, fixed subintervals
of length $\delta t$, and $k$ -- the number of counts 
in such an interval -- is Poisson distributed: 
\begin{equation}
P( k | PCP, \Lambda ) = 
{\Lambda ^{k} e^{-\Lambda } 
\over 
k!} ,
\label{poisson}
\end{equation}
\noindent
with parameter
\begin{equation}
\Lambda \equiv \lambda \delta t .
\label{poisson_parameter}
\end{equation}
\noindent
[Note: the count rate is expressed either
per unit time, with $\lambda$ (dimension is $s^{-1}$),
or per interval with $\Lambda$ (dimensionless).]
Throughout, we assume that the arrival of a photon in any 
interval is independent of that in any other 
non overlapping interval; {\it i.e.}, 
the joint probability distribution of the random variables 
describing photon arrival in the two disjoint intervals
is the product of the individual distributions.
(Do not confuse the photon detection process with 
the possibly random process describing the source
intensity as a function of time -- which 
is typically correlated from one time to the next.
See page 99 of \cite{brill_2} for a discussion of this
issue, known as {\it doubly stochastic processes}.)
We make considerable use of the fact 
that an event probability in an interval 
is the product of the probabilities in its subintervals.

%---------------------------------------------------
\subsubsection{Time-Tagged Event (TTE) Data}
\label{TTE}
%---------------------------------------------------

The recording mode called {\it event} or {\it time-tagged} data 
is common in X-ray and $\gamma$-ray astronomy,
and capable of the highest time resolution. 
In this mode the detection times of individual photons are 
recorded.
In principle, the raw data consist of a set of $N$  
{\it photon arrival times}
\begin{equation}
D_{TTE}: \{ t_{n}, \ \  n = 1, 2, 3, \dots , N \} 
\label{tte_data_1}
\end{equation}
\noindent
over the range of times during which the instrument was active.
See \cite{brill_2} for a discussion of this kind of process, 
consisting of discrete events -- called {\it point processes} 
in the statistics literature.

In practice, of course, these times are recorded 
with small but finite resolution -- 
the photons are assigned to narrow bins, 
as described in connection with equation (\ref{poisson}).
However, in most data systems there are two 
reasons for not thinking of these as ordinary bins:
First, the time interval is very short
(for BATSE $\delta t =2\mu$-sec) compared to 
time scales of astrophysical interest. 
Second, the actual number of photons in the interval 
is not recorded -- just whether one or more photons
have arrived.\footnote{
On the other hand, some systems (including BATSE and RXTE) 
have several detectors operating essentially independently and simultaneously, 
and photons from different detectors can be recorded
with the same time stamp.  I ignore these complications.} 
These considerations justify our thinking of this analysis
as {\it bin free} and calling the 
intervals ``ticks,'' by analogy to 
a digital clock, instead of bins.

We introduce an integer time index $m$ through
\begin{equation}
t_{m} = m \ \delta t,
\label{time_index}
\end{equation}
\noindent
where for an observation of duration $T = M \delta t,
m = 1, 2, 3, \dots , M$.
The data consist of a set of $N$ indices, one for each photon:
\begin{equation}
D_{TTE}: \{ m_{n}, \ \ n = 1, 2, 3, \dots , N \},
\label{tte_data_2}
\end{equation}
\noindent
meaning that photon $n$ is 
detected at time $ m_{n} \ \delta t$.

A third way to represent these data,
fully equivalent to the two above, 
is in terms of the observable $X_{m}$ defined by
\begin{equation}
D_{TTE}: X_{m} = \left\{ \begin{array}{cc}
                0 & \mbox{no photons during tick}\ m \\
                1 & \mbox{otherwise}
               \end{array}
                   \right.
\label{tte_data_3}
\end{equation}
The probabilities of these values are  
\begin{equation}
\begin{array}{ll}
P( X_{m} = 0 | \Lambda) &= p_{0} \equiv e^{- \lambda \delta t} = e^{-\Lambda}\cr
P( X_{m} = 1 | \Lambda)  &= p_{1} \equiv 1 - p_{0}\cr
\end{array}
\label{element}
\end{equation}
\noindent
Strictly speaking $p_{1}$, 
since it is the probability of one {\it or more} photons,
is not proportional to the Poisson rate parameter.
However, since this parameter is small -- 
typically $\approx 0.01$ counts per tick, or less -- we have
\begin{equation}
p_{1} = 1 - e^{- \lambda \delta t} \approx \lambda \delta t \equiv \Lambda
\label{approx_p}
\end{equation}
\noindent
This approximation is useful at a few points, 
but the main analysis does not depend on it.
Technically the above conditions define a 
{\it finite Bernoulli lattice process}
\cite{stoyan}, since $X$ takes on one of two possible 
values over a finite range of discrete times.
Here I nevertheless follow common usage in 
referring to this as a Poisson process.

By the independence assumption discussed above,
the joint probability of all the events $X_{m}$ is just 
the product of the probabilities of the individual events.
That is to say, defining {\MyScript M}$_{1}(\Lambda, T)$
as the Poisson process over interval $T$, with
rate $\Lambda$ per tick, we have
\begin{equation}
P[ D_{TTE} | \mbox{\MyScript M}_{1} (\Lambda, T) ] = 
\prod_{m=1}^{M} P( X_{m} | \Lambda ) =
p_{1} ^{N} (1 - p_{1})^{M-N} 
\label{p_lambda}
\end{equation}
\noindent
since $N$ ticks contain a photon and the remaining $M-N$ do not.
This probability is maximized at 
$p_{1} = {N \over M}$, 
and equation (\ref{element}) gives as the most probable rate
\begin{equation}
\lambda = - { 1 \over \delta t } log( 1 - {N \over M} ),
\label{exact_rate}
\end{equation}
\noindent
which in the approximation of equation (\ref{approx_p}) 
reduces to 
\begin{equation}
\lambda = { 1 \over \delta t } {N \over M}.
\label{rate_tte}
\end{equation}
\noindent

In view of the form of equation (\ref{p_lambda})
we now switch from $\Lambda$ to $p_{1}$ as the model parameter, 
to simplify the analysis. 
Furthermore, this change motivates selection of 
the following prior distribution:
\begin{equation}
P( p_{1} | \mbox{\MyScript M}_{1} ) =  \left\{ \begin{array}{cc}
                            1 & \mbox{for $0 \le p_{1} \le 1$ } \\
                            0 & \mbox{otherwise }
                                    \end{array}
                            \right.
\label{prior_tte}
\end{equation} 
\noindent
This normalized prior  
[$\int P( p_{1} | \mbox{\MyScript M}_{1} ) dp_{1} = 1$] 
assigns probability uniformly to all physically realizable values.
It is therefore less arbitrary than 
some priors adopted in Bayesian statistics,
and we adopt it here in preference to 
alternatives, such as uniform in $\Lambda$ or
with cutoffs corresponding to some sort of
{\it a priori} upper or lower limits on 
counting rates.

To evaluate the global likelihood in equation (\ref{global_like_1}),
multiply the likelihood in (\ref{p_lambda})
by the above prior and integrate 
\begin{equation}
\int P[ D_{TTE} | \mbox{\MyScript M}_{1}(p_{1})] P( p_{1} | \mbox{\MyScript M}_{1} ) dp_{1}
= \int _{0}^{1} p_{1}^{N} ( 1 - p_{1} ) ^{M-N} dp_{1} 
= B( N+1, M - N + 1) 
\label{beta}
\end{equation}
\noindent
where the {\it beta function} $B$ can be written in terms
of the gamma function [\cite{jeffrey}, \S 11.1.7]: 
\begin{equation}
B(x,y) = {\Gamma (x) \Gamma (y) \over \Gamma( x + y ) } 
\label{beta_gamma}
\end{equation}
\noindent
In summary, the global likelihood for the single rate model  
is this simple function: 
\begin{equation}
\mbox{\MyScript L}( \mbox{\MyScript M}_{1} | D_{TTE} ) = {\Gamma (N+1) \Gamma (M-N+1) \over \Gamma(M+2) }
= {N! (M-N)! \over (M+1)!}.
\label{final_tte}
\end{equation}
\noindent
It may seem peculiar that  
this likelihood for a constant rate depends 
not at all on the distribution of the photon 
times within the interval -- but 
on only the length of the interval and 
the number of photons in it.
This quantity measures the likelihood 
of a single-rate model only when 
compared with the analogous quantity 
for another model class.
This relationship is detailed 
in \S \ref{determine_model} where
a single-rate, unsegmented model 
is compared with 
a two-rate, segmented model 
for the same data.

Note: had we used the probabilities from the 
truncated Poisson distribution --
$e^{-\Lambda}$
and 
$\Lambda e^{-\Lambda}$
for zero and one photon,
respectively --
we would have arrived at 
\begin{equation}
\mbox{\MyScript L}
( \mbox{\MyScript M}_{1} | D_{TTE} ) = {\Gamma (N+1) \over (M+1)^{N+1} },
\label{not_norm}
\end{equation}
\noindent
a result obtained by \cite{raftery_akman} -- and applied to 
a study of the intervals between coal-mining disasters -- but
with a prior somewhat different from ours.
In fact, equations (\ref{final_tte}) and (\ref{not_norm})
give very similar values, 
which may be taken as evidence that 
details of the prior do not matter very much.
Equation (\ref{final_tte}) will be used here.

%---------------------------------------------------
\subsubsection{Binned Data}
%---------------------------------------------------

Sometimes the data are pre-binned 
% aboard the spacecraft 
into $M$  evenly spaced intervals:
\begin{equation}
D_{BIN}: \{ X_{m}, \hskip .1in m = 1, 2, \dots, M\}, 
\label{bin_data}
\end{equation}
\noindent
where the integer $X_{m}$ is the number of photons
detected during the $m$-th such time interval.
Taking the rate per bin to be constant,
say $\Lambda$, 
the counts in a given bin obey Poisson statistics
for this rate:
\begin{equation}
P( X_{m} | \Lambda ) = 
{\Lambda ^{X_m} e^{-\Lambda} 
\over 
X_{m}!}
\end{equation}
\noindent
Independence of the counts $X_{m}$
yields for the likelihood: 
\begin{equation}
P[ D_{BIN} | \mbox{\MyScript{M}}_{1}(\Lambda )] = 
\prod _{m=1} ^{M} {\Lambda ^{X_m} e^{-\Lambda} \over X_{m} !} =
{\Lambda ^{N} e^{- M\Lambda}
\over
\prod_{m=1}^{M} X{_m} ! }
\label{likelihood}
\end{equation}
\noindent
where $N = \sum_{m=1}^{M} X_{m}$ is 
the total number of photons.
The maximum of this
probability occurs at the same value given in equation (\ref{rate_tte}).

Note that the denominator in equation (\ref{likelihood})
has the property that its value for an interval is just the product of
its value for two or more subintervals.  Hence this factor
cancels out in a comparison of segmented {\it vs.} unsegmented 
versions of a given model, 
and we omit it.
With $\Lambda$ as the parameter, we adopt the nonuniform but normalized prior 
\begin{equation}
P( \Lambda | \mbox{\MyScript M}_{1} ) =  \left\{ \begin{array}{ll}
                            (1 - e^{-C} ) e^{- \Lambda} & \mbox{$0 \le \Lambda \le C$ } \\
                            0 & \mbox{$\Lambda < 0$ or  $\Lambda > C$ }
                                    \end{array}
                            \right.
\label{prior_bin}
\end{equation} 
\noindent
This prior, while nonuniform in $\Lambda$, 
corresponds to the same uniform $p_{0}$-distribution  
used in the TTE case.
It is a special
case of the Gamma distribution (power law times exponential) commonly
used in Bayesian inference with the Poisson distribution 
[\cite{o_hagan,lee}].  This particular form 
reflects the prior belief that the rate 
is unlikely to exceed a specific, if approximate, value
set by instrumental considerations (which in 
turn may be guided in the instrument design phase by  
the maximum expected source brightness).  
For example, $C$ might be reckoned as roughly 
the bin interval divided by the instrument deadtime.

Integrating the above likelihood times this prior,
and -- absent a preferred value of $C$ -- 
taking the limit $C \rightarrow \infty$ ({\it i.e.} 
allowing bin counts to have any positive value) gives:
\begin{equation}
\mbox{\MyScript L}( \mbox{\MyScript M}_{1} | D_{BIN} ) = 
\int_{0}^{1} \Lambda^{N} e^{-(M+1)\Lambda } d\Lambda  =
{\Gamma(N+1) \over (M+1)^{N+1}}, 
\label{final_bin}
\end{equation}
\noindent
curiously identical to equation (\ref{not_norm}).

%---------------------------------------------------
\subsubsection{Time-to-Spill Data}
%---------------------------------------------------

The last data mode considered is called {\it time-to-spill} (TTS).
To reduce the telemetry data rate, 
only every $S$-th photon is recorded,
where $S$ is an integer\footnote{The data-descriptive constant $S$ 
is not to be confused with a model parameter.}
(typically 64 for the BATSE TTS mode):
\begin{equation}
D_{TTS}: \{ \tau_{n}, \ \  n = 1, 2, \dots, N-1\}, 
\label{tts_data}
\end{equation}
\noindent
where $\tau_{n}$ is the interval 
between the $n$-th and the $n+1$-th spill events.
It is well known that the distribution of 
such intervals is given by 
the gamma, or Erlang, distribution [\cite{billingsley,haight}]:
\begin{equation}
P( \tau _{n} |  \Lambda) = 
{ \Lambda ^{S} \over \Gamma (S) }
\tau_{n}^{S-1} e^{ - \Lambda \tau_{n}} 
\label{beta_distribution}
\end{equation}
\noindent
The usual independence assumption yields:
\begin{equation}
P[ D_{TTS} | \mbox{\MyScript M}_{1}( \Lambda ) ] = 
[{\Lambda^{S} \over \Gamma(S)} ]^{N-1} (\prod_{n=1}^{N-1} \tau_{n} )^{S-1} e^{- \Lambda M },
\label{likelihood_tts}
\end{equation}
\noindent
where $M = \sum_{n=1}^{N-1} \tau_{n}$ is the total length 
of the interval.
As expected, this probability 
is maximum at
\begin{equation}
\Lambda = {S N \over M }.
\label{rate_tts}
\end{equation}
\noindent
Equation (\ref{likelihood_tts}),
integrated  
with the same prior in equation (\ref{prior_bin}),
and again taking the limit $C \rightarrow \infty$,
gives 
\begin{equation}
\mbox{\MyScript L}( \mbox{\MyScript M}_{1} | D_{TTS} ) = 
{ 
( \prod_{n=1}^{N-1} \tau_{n} ) ^ {S-1}
\over 
\Gamma(S)^{N-1} 
}
{\Gamma[ S(N-1)+1] \over (M+1)^{S(N-1)+1}} .
\label{final_tts}
\end{equation}
\noindent
Note that the interpretation of $\tau$ 
in terms of the true photon rate 
involves the same issue raised in the TTE case: 
because of detector dead-time, accumulation of
$S$ detector counts occurs at a slightly
lower rate than does arrival of $S$ photons.
In practice the corresponding corrections
can usually be ignored [{\it cf.} equations (\ref{exact_rate}) 
and (\ref{rate_tte})].

%----------------------------------------
\subsection{Evidence for a Segmented Poisson Rate Model}
\label{evidence_segmented}
%----------------------------------------

The previous section yielded estimates of 
the relative probabilities of the simplest  
model -- namely the single constant-rate Poisson 
{\MyScript M}$_{1}(\Lambda)$  -- 
for TTE, binned, and TTS data in equations (\ref{final_tte}),
(\ref{final_bin})
and 
(\ref{final_tts}), respectively.
These global likelihoods  
depend on only 
$N$ and $M$, so we denote them as
\begin{equation}
\mbox{\MyScript L}( \mbox{\MyScript M}_{1} | D ) = \phi_{D}(N,M)
\label{phi_def}
\end{equation}
\noindent
where $D$ here denotes the datatype, and
where 
\begin{equation}
\phi_{TTE}(N,M) = {\Gamma (N+1) \Gamma (M-N+1) \over \Gamma(M+2) }
\label{phi_tte}
\end{equation}
\noindent
\begin{equation}
\phi_{BIN}(N,M) = {\Gamma(N+1) \over (M+1)^{N+1}}
\label{phi_bin}
\end{equation}
\noindent
and
\begin{equation}
\phi_{TTS}(N,M) = { 
( \prod_{n=1}^{N-1} \tau_{n} ) ^ {S-1}
\over 
\Gamma(S)^{N-1} 
}
{\Gamma[ S(N-1)+1] \over M^{S(N-1)+1}}
\label{phi_tts}
\end{equation}

These results will now be used 
to estimate the model
in which the observation interval 
is broken into two sub-intervals 
over which the rates are assumed to be constant but different.
({\it Cf.} the example at the beginning of \S  \ref{compare}.)  
In the statistics literature, the point separating such 
segments is called a {\it changepoint} in the time series,
because the underlying process changes abruptly there.
Denote the two-segment model with constant Poisson rates 
{\MyScript M}$_{2}(\Lambda_{1}, \Lambda_{2}, t_{cp})$,
where $t_{cp}$ denotes the {\it changepoint} -- {\it i.e.}, the 
time at which the rate switches from $\Lambda_{1}$ to $\Lambda_{2}$.
In the notation of \S \ref{compare},
the full interval $T$ is partitioned 
into two intervals, $T_{1}$ and $T_{2}$, containing the times
less than and greater than $t_{cp}$, respectively.

The probability of the compound model is, 
by the same independence assumption discussed above, 
just the product of the probabilities of the two 
segments considered separately:
\begin{equation}
P[ D(T) | \mbox{\MyScript M}_{2} (\Lambda_{1}, \Lambda_{2}, t_{cp}) ]
= 
P[ D_{1} | \mbox{\MyScript M}_{1}(\Lambda_{1}, T_{1} ) ]
P[ D_{2} | \mbox{\MyScript M}_{1}(\Lambda_{2}, T_{2} ) ]
\end{equation}
\noindent
where $D_{1}$ is the data in interval $1$, {\it etc}.
Thus the global likelihood for the two-rate model is
\begin{equation}
\mbox{\MyScript L}( \mbox{\MyScript M}_{2} | D ) = 
\int dt_{cp}
\int d\Lambda_{1}
\int d\Lambda_{2}
P_{cp}(t_{cp})
P[D_{1} | \mbox{\MyScript M}_{1}(\Lambda_{1}, T_{1} )] P_{\Lambda}(\Lambda_{1} ) 
P[D_{2} | \mbox{\MyScript M}_{1}(\Lambda_{2}, T_{2} )] P_{\Lambda}(\Lambda_{2} ) ,
\label{cp_global_like}
\end{equation}
\noindent
where $P_{\Lambda}$ is the rate prior appropriate to the data type,
and $P_{cp}$ is the changepoint time prior.
Note that the variables
$\Lambda_{1}$, 
$\Lambda_{2}$, and
$t_{cp}$,
which are essentially nuisance parameters here,
are integrated out, and the likelihood is therefore independent of them.

Consider now the TTE case.
For actual data, time
is discrete ({\it cf}. \S \ref{TTE}),
so the integral in equation (\ref{cp_global_like}) is a sum 
and we denote the changepoint location 
by the integer $m_{cp}$.
One could consider jumps  
at arbitrary clock times, $m$,
but it simplifies the procedure to 
test for possible changepoints only at the arrival 
of an actual photon.  
Thus we parametrize the changepoint as 
\begin{equation}
m_{cp} = m_{n_{cp}}
\label{change_point}
\end{equation}
\noindent
for some photon index $n_{cp}$.
This simplification 
merely ignores the difference between points that 
identically divide the photons.
Further, after carrying out the two
$\Lambda$ integrals, we can write  
the $t_{cp}$-integrand (or rather the
corresponding discrete time summand) in
equation (\ref{cp_global_like}) 
as a simple function of 
only the changepoint index $n_{cp}$,
through the relations
\begin{equation}
N_1 = n_{cp},
\end{equation}
\noindent
\begin{equation}
N_2 = N - N_{1} = N - n_{cp},
\end{equation}
\noindent
\begin{equation}
M_1 = m_{n_{cp}},
\end{equation}
\noindent
and
\begin{equation}
M_2 = M - M_{1} = M - m_{n_{cp}},
\end{equation}
\noindent
From the expressions above, and with the definition
\begin{equation}
\psi(n_{cp}) = \phi(N_{1},M_{1}) \phi( N_{2}, M_{2} ), 
\label{def_psi}
\end{equation}
\noindent
we have  
\begin{equation}
\mbox{\MyScript L}( \mbox{\MyScript M}_{2} | D )
= \sum_{n_{cp}} \psi(n_{cp}) \Delta t_{n_{cp}} 
\label{cp_global_like_1}
\end{equation}
\noindent
where the factor $\Delta t_{n_{cp}}$ -- defined as the 
time interval between successive photons --
corresponds to a prior uniform in $m$, even though 
the sum in this equation is over $n_{cp}$, not $m$.
In fact, the code in Appendix A omits this factor,
because it appears to be a small correction in 
all the cases studied so far.
The changepoint parameterization is slightly different
for the other data modes; details are omitted.

%----------------------------------------
\subsection{Deciding Between Segmented and Unsegmented Model}
\label{determine_model}
%----------------------------------------

The idea now is simple: 
compare the $J( \mbox{\MyScript M}_{k}, D )$ values 
from equation (\ref{total_like}) of 
the unsegmented, single rate model $\mbox{\MyScript M}_{1}$ and 
the segmented, two-rate model $\mbox{\MyScript M}_{2}$,
in terms of the odds ratio:
\begin{equation}
\mbox{\MyScript O}_{21} = 
{ {\mbox J}( \mbox{\MyScript M}_{2}, D ) 
\over
\mbox{ J}( \mbox{\MyScript M}_{1}, D ) }
\label{final_odds}
\end{equation}
\noindent
This ratio, with the prior odds ratio equal to one,
is often called the {\it Bayes factor}.
If this ratio favors a segmented model,
it is straightforward to compute from equation (\ref{def_psi}) 
the most probable changepoint location 
from among all possible changepoints.
Finally and almost trivially, equation (\ref{exact_rate}) 
or equation (\ref{rate_tte})
determines the corresponding rates.
The appendices contain computer code for
all the necessary computations, 
and the procedure 
is demonstrated on real data 
in \S \ref{example_batse}  below.

%----------------------------------------
\subsection{Multiple Change Points}
\label{blocks}
%----------------------------------------

As discussed earlier, 
the overall goal is to find the optimum block decomposition
of the data -- {\it i.e.} into a piecewise constant
representation.
The rigorously correct way to do this would be as follows.
Let an arbitrary number, say $N_{cp}$, 
of changepoints divide the observation interval
into $N_{cp}+1$ subintervals.
Compute the global likelihood, $\mbox{\MyScript L}(N_{cp})$
of this multiple changepoint decomposition;
the value of $N_{cp}$ which maximizes this quantity 
is the most probable number of
changepoints.  It is then a simple matter to 
find the most probable locations of the change 
points themselves, and the most probable values 
of the rates for each of the corresponding segments.

The case $N_{cp} = 2$ is relatively easy.  
In fact, the corresponding global likelihood -- a 
function of the two changepoint location indices --
can be computed with matrix operations that are
quite efficient in MatLab.
Some thinning of the data is necessary for the 
cases in which the number of photons is so 
large that the corresponding $N \times N$ 
matrix is too big.
However, for more than 2 changepoints 
direct computation 
quickly becomes impractical.
Therefore a simple iterative procedure was adopted 
as an attempt to approximate multiple
changepoint determination.
Start with the whole observation interval.
Use the above method to decide between 
not segmenting this interval 
and segmenting it, with one changepoint, 
into two subintervals.
If the latter is favored by the Bayes factor,
apply the same procedure to both of the
resulting subintervals.
Continue in the same vein, applying
the procedure to all new subintervals
created at the previous step.
That this method works approximately, 
but not exactly, is indicated
by the fact that an
algorithm that handles two simultaneous change-points
({\it i.e.} $N_{cp}$ = 2) 
gave results similar, but not identical,
to those obtained with iterative 
application of the single changepoint algorithm.

What stops this iteration?
The obvious halting condition is that 
the odds ratios favor unsegmented 
models for all subintervals.
Unfortunately this is too simple in practice.
In the analysis of large data sets 
there are typically many computed odds ratios which are
greater than $1$ by only a small amount.
Decisions based on these ``coin flips'' are wrong 
about half the time, 
subdividing many intervals that shouldn't be.

Since these cases tend to be short 
intervals containing only a few photons, 
much of the problem can be swept under the
rug by imposing a minimum number of 
photons allowed in subintervals.
A second approach is to 
impose a prior odds ratio that disfavors 
segmenting  -- {\it i.e.}, is biased toward 
keeping intervals unsegmented unless the
odds ratio is strongly in favor of segmenting.
(There is a simple argument in support of this
second idea:
an overall statistical assessment should take
into account the number of roughly independent experiments 
carried out; this is on the order of the 
largest reasonable number of segmentation points, 
which in turn is determined by the resolution 
of the observation interval.
This leads to a prior ratio 
in equation (\ref{prior_odds}) of
$\rho \approx { \mbox{length of data interval} \over \mbox{desired time resolution} }$.
It also has the advantage that it 
avoids the other idea's bias against short intervals.
Unfortunately, this argument 
probably cannot be justified  
within the Bayesian formalism.
Nevertheless, numerical experiments support 
the use of one or the other of these ideas.)
The best approach may be to combine
both, as was done by 
\cite{bernaola} in a similar segmentation algorithm, 
based on the {\it Jensen-Shannon divergence measure} 
in place of the likelihood, and applied to 
automatic detection of structure in DNA sequences.
\cite{gustafsson_2} uses a
{\it stopping rule} 
based on somewhat different considerations.
The code in Appendix B
shows one way to carry out iterative 
segmentation and such a composite halting logic.

%========================================
\section{BATSE $\gamma$-ray burst data}
\label{example_batse}
%----------------------------------------

This section demonstrates the 
method just described by applying it
to $\gamma$-ray data from BATSE.
The basic algorithm
is employed to determine the detailed structure of 
pulses, such as are known to make up 
the time-profiles of many $\gamma$-ray bursts
[\cite{attributes,attributes_2}].

\begin{figure}[htb]
\hspace{6.5in}
\vspace{3.5in}
\par
\hskip -.5in
\special{bb_fig_1.epsf}
\caption{Changepoint location in BATSE data for Burst Trigger 0551.
(a) Binned counts for comparison: 100 time bins,
of width 9.42 ms.  
(b) For TTE data: $log_{10}$ of the odds ratio in favor of segmentation,
as a function of the changepoint location.
(c) Same for binned data.
(d) Same for TTS data.
Vertical lines in all panels are at the maximum odds ratio;
in (a) those for TTE and TTS modes are indistinguishable 
and appear as a solid line.
}
\end{figure}

Figure 1 depicts the logarithms of the odds ratios
as a function of the position of the changepoint 
for BATSE data from the burst 
denoted Trigger 0551.
The top panel shows for comparison the binned counts 
as a function of time (in microseconds).
The raw data comprises about $29,000$ photons.
On the same time axis, the other
panels show the 
logarithms of the odds
ratio in equation (\ref{final_odds}),
for TTE, binned, and TTS data, in order,
as a function of the location of the changepoint.
The binned and TTS data are derived directly
from the TTE data.
The spill data was constructed simply by
sampling every $64$-th photon from the TTE data.

Note several things:
(1) The actual odds ratios are all astronomically
large in favor of segmentation.
(2) The most probable changepoint location is
indicated with vertical dotted, dashed, and dot-dashed lines.
If the actual odds ratios were plotted,
this would be an extremely sharp maximum,
indicating that there is very little 
uncertainty in the changepoint location.
(3) The TTE and TTS changepoints 
are very nearly equal -- suggesting
that this method is rather efficient 
at extracting information 
from TTS data, and also 
that little information is lost in this mode.
The fact that the value for the binned 
data is slightly different is consistent
with the expected loss of time-resolution 
entailed by binning.

\begin{figure}[htb]
\hspace{4in}
\vspace{4.5in}
\par
\hskip -.05in
\special{bb_fig_2.epsf}
\caption{Bayesian Blocks
for the same data as in Figure 1,
determined as explained in the text.
(a) TTE data; 
(b) TTS data;
(c) binned data}
\end{figure}

Figure 2 shows the result of iterating the 
segmentation procedure on the same TTE data.
The Bayesian blocks are indicated with solid lines.
The vertical dotted lines are the locations 
of pulses determined by a simple pulse finding routine 
that basically selects statistically significant local maxima;
this algorithm will be described in \cite{attributes_2}.

One can derive properties of the pulses from 
this block representation.
In a separate paper [\cite{attributes_2}], 
this method will be used to
determine peak times, amplitudes, and rise and
fall times for gamma ray bursts.
Specifically, we use the Bayesian Blocks
technique to make crude estimates of 
the locations, amplitudes, and widths of 
the pulse structures within a burst,
without a parametric pulse model
and dealing with pulse overlap in a trivial way.
The peak time and amplitude are taken as the 
center and height of the highest block in the pulse, and
exponential rise and decay times are estimated 
using a simple quadrature of the corresponding halves
of the burst profile.
Then these crude pulses are used as initial guesses 
for a numerical routine that truly deconvolves 
overlapping pulses by fitting a parametric model.
The initial guess is very important for the
convergence of this fitting procedure to the
(hopefully) global optimum;
results with the Bayesian Blocks have proven 
very satisfactory.
The lowest block provides a good estimate
of the constant post-burst background, and will do so 
as long as the burst ends before the observation 
terminates.

%========================================
\section{Conclusions}  
%----------------------------------------

The method developed here is 
applicable to all the photon event data modes
common in high energy astrophysics: 
time-tagged events, binned counts,
and time-to-spill data.
The fundamental element of the method is 
a way to decide whether 
a single Poisson rate or 
two different rates
is the better model for an interval.
This decision is applied iteratively 
to build up a piecewise constant  
model of the data.
This analysis method imposes no lower limit 
on the time scale;
any such limits are set by the
the data themselves.

The Bayesian Blocks method is designed to extract localized 
signals from counting data where statistical fluctuations
are important. 
It is probably not useful in situations
that require lots of time-averaging to extract 
coherent, global signals such as periodic 
or quasi-periodic variations.

Future work will include investigation 
of ways to determine multiple changepoints
more rigorously.
The principles behind a maximum likelihood 
determination of the number and location
of changepoints is straightforward, 
and can surely be made computationally 
feasible.
I have recently become aware of 
work by \cite{chib}
developing a Markov chain Monte Carlo
procedure for Bayesian estimation of 
multiple changepoint models that may 
be applicable to this problem. 
\cite{phillips_smith}
may also be of relevance.

In addition, it
will be useful to extend the methods
given here to include variable rates
across the blocks, or other
departures from the constant-rate model.
I have explored both linear
and exponentially varying rates.
The approach in \cite{sugiura_ogden} 
may be useful for this problem.
I am pursuing extensions of the 
basic idea underlying Bayesian Blocks 
to higher dimensions;
in particular spatial structure 
can be elucidated, and backgrounds 
removed, from two-dimensional 
photon counting data with generalizations 
of the one-dimensional algorithms 
given here.

It is also relatively easy 
to extend this methodology to 
a multivariate context -- determination 
of block structure in pairs of time series 
in which it is assumed that the segmentation 
points occur at the same times in the 
two data series; of course, the rates 
are not in general the same.
This will be particularly useful 
for BATSE gamma-ray burst data 
that consists of simultaneous photon counting in 
four broad energy channels.
In this context, it will be useful to allow for, {\it e.g.} 
the known fact that there are time delays in the
burst structures as a function of photon energy.
Similarly, known gaps during which the instrument is
not sensitive can be readily handled.

What to do with Bayesian Blocks?
This depends on the context.  
For the pulse problem in $\gamma$-ray burst work
(\S \ref{example_batse})
we have indicated the use of the blocks to determine 
pulse attributes, at least in a crude way, without
the need to adopt a specific model for pulse shapes.
These attributes can in turn be used 
as starting guesses for further, parametric,
nonlinear optimization, as discussed above.
It is expected that many different uses can
be made of Bayesian block decomposition.

Work is in progress in collaboration with 
Paul Hertz, 
Elliott Bloom,
Jay Norris and
Kent Wood,
to use Bayesian Blocks 
to determine whether short-time-scale
structure, or {\it bursts}, 
are present in Cygnus X-1.
There is a long debate in the literature about 
the reality and meaning of short (millisecond)
bursts in this accretion system.
Almost certainly our approach will
either detect or place
upper limits on bursts, and has the 
possibility of detecting individual bursts 
at a high significance level.
A different approach to this same problem,
also using a Bayesian framework, 
was presented at a recent 
meeting of the High Energy Astrophysics 
Division of the American Astronomical 
Society [\cite{marsden_rothschild}].

Note added in press:  For TTE data,
consider the time-scale transformation
$\delta t \rightarrow {1 \over \alpha} \delta t,
M \rightarrow \alpha M$, for $\alpha$ any integer $> 1$.
This amounts to refining the clock ticks but
leaving the photon times unchanged.
Under this transformation the estimated block structure must 
be unaltered: the changepoint times and photon rates will stay fixed
(although of course the rates {\it per tick} will decrease by
a factor of $\alpha$).
By considering arbitrarily large $\alpha$ it follows that
the asymptotic form (for $M \rightarrow \infty$) 
of equation (\ref{final_tte}) 
can be used without appreciable error.
Details of this simplification 
will be posted on the World Wide Web site 
referenced in the Appendix and,
together with a solution of the multiple changepoint problem 
using Markov Chain Monte Carlo methods, 
will be the subject of a future paper.

\acknowledgments

I am especially grateful to Tom Loredo 
for encouragement, numerous technical suggestions,
and careful reading of
several versions of the manuscript.
Thanks also to Jay Norris and Jerry Bonnell for 
initiating the gamma-ray burst research 
that led to this work and for
numerous useful suggestions, and to 
Eric Kolaczyk,
Iain Johnstone, 
and
Peter Cheeseman
for statistical advice.
I thank Bob Hogan and Mark Showalter,
plus members of the SLAC Astrogravity group --
Elliott Bloom,
Chris Chaput,
Daniel Engovatov,
Gary Godfrey, 
Andrew Lee,
and
Ganya Shabad -- 
for helpful comments 
and assistance. 
I am grateful to 
David Marsden and Rick Rothschild for an advance
copy of their paper,
and Bill Fitzgerald and Fredrik Gustafsson
for helpful comments.
This work is supported by grants from NASA's
Astrophysics Data Program,
the Compton Gamma-ray Observatory 
Guest Investigator Program, 
and the NASA-Ames Director's Discretionary Fund. 
The NASA data shown are from 
the BATSE instrument on the 
Compton Gamma Ray Observatory.

\newpage
%========================================
\section{Appendix}  
%----------------------------------------

This appendix contains 
MatLab\footnote{\copyright the Mathworks, Inc.}
code,
an array-based data processing system.
These MatLab scripts, and sample data allowing 
the reader to reproduce Figure 2 of this paper, can be found on 
the World Wide Web at: \verb+http://george.arc.nasa.gov/~scargle/papers.html+.
(See \cite{buck_donoho}
for a description of the philosophy of
publishing scientific research in such a 
way the the reader can reproduce all results.)
Much of this code is similar to that 
of IDL\footnote{\copyright Research Systems, Inc.} 
and other similar software packages for 
data analysis, and
can be considered as pseudocode for the procedure.

A few comments about the MatLab syntax 
is in order.

The function line at the beginning of each
routine identifies the input and output
variables.  It will be seen that multiple 
input and output variables are possible;
and the input and output variables are 
arrays (matrices, vectors, or scalars) in general.

The symbols $*$ and $/$ specify 
matrix multiplication and division, respectively.
Overriding the matrix operation in favor of
the simple term-by-term operation
is indicated by a dot ($.$) before $*$ or $/$.
The statement \verb$[ a_max, i_max ] = max( x )$,
where \verb$x$ is a vector, returns both the value
of the maximum of the array, and the index, \verb$i_max$
at which this maximum is achieved.
The function \verb$gammaln$ is a built-in
function that evaluates the natural logarithm
of the gamma function of the argument array.

On any line, everything following the symbol \verb$%$ 
is treated as a comment and not processed.
Three dots ($\dots$) at the end of a line indicates
continuation onto the next line.

The command \verb+find+ returns the indices
of its argument that satisfy the condition
specified in the argument.
\verb+isempty+ is a logical function to 
determine whether the argument has been defined yet.
\verb+reverse+ simply reflects an array,
and \verb+ceil+ and \verb+floor+ are 
rounding of a real number to the next highest and lowest 
integer, respectively.

The expression \verb+a'+ means the matrix transpose of \verb+a+.

\subsection{Appendix A: Find A Change Point}

This Appendix gives MatLab code for
the procedure to find a single change 
point, as
described in \S  \ref{determine_model} of the text.
The computation is particularly efficient because
the evaluation of the global likelihoods can be 
carried out completely in terms of 
array operations on the 
vector containing all the candidate 
changepoints.

\SS
\begin{verbatim}
function [ change_point, odds_21, log_prob, log_prob_noseg ] = ...
    find_change( photon_times, t_0, t_n )
%
% Find most probable two-rate model for Poisson arrival time data, 
% based on Bayesian analysis.
%
% Input:  photon_times -- photon arrival times
%                         (Note: These must be microseconds, not seconds,
%                         because the time values correspond to the 
%                         clock rate at which the data are sampled.)
%         t_0          -- time just previous to photon_times(1)
%         t_n          -- time just after last time in photon_times
%
% Output: change_point -- index of "photon_times" which provides the maximum 
%                         likelihood segmented model (that is, with one   
%                         Poisson rate to the left of 
%                               photon_times(change_point) 
%                         and another to the right
%              odds_21 -- odds ratio: 2 unequal rates / 1 rate
%             log_prob -- log probability of segmented model, as a  
%                         function of changepoint 
%       log_prob_noseg -- log prob of nonsegmented model
%--------------------------------------------------------------------------------------

dt_half = 0.5 * diff( photon_times );
n_ph = length( photon_times ); % Number of photons

min_time = photon_times(    1 );
max_time = photon_times( n_ph );

t_left  = 0.5 * ( t_0 + min_time );
t_right = 0.5 * ( max_time + t_n );

% Number of microsecond "ticks" in the whole (extended) interval:
n_ticks = t_right - t_left + 1; 

%------------------------------------------------------------------
%    Evaluate log-probability of the unsegmented model:
%------------------------------------------------------------------

log_prob_noseg = gammaln( n_ph + 1  ) + ...
                 gammaln( n_ticks - n_ph + 1 ) - ...
                 gammaln( n_ticks + 2);

%------------------------------------------------------------------
% Evaluate the log-probability of the segmented model as a 
% function of changepoint; find optimum changepoint.
%------------------------------------------------------------------

% Array of trial changepoints:
n_1  = (1: n_ph - 1)';
n_2 = n_ph - n_1;

m_1 = photon_times( n_1 ) + dt_half( n_1 ) - t_left;
m_2 = n_ticks - m_1;

log_prob = - 1.e55 * ones( n_ph, 1 ); % mark all points as invalid
arg_1 = m_1 - n_1 + 1;
arg_2 = m_2 - n_2 + 1;
ii = find( arg_1 > 0 & arg_2 > 0 ); % indices of valid points
log_prob(ii) = gammaln( n_1(ii)+1 ) + gammaln( arg_1(ii) ) - gammaln( m_1(ii)+2 );
log_prob(ii) = log_prob(ii) + gammaln( n_2(ii)+1 ) + gammaln( arg_2(ii) ) - ...
              gammaln( m_2(ii) + 2 );
[ max_log, change_point ] = max( log_prob(ii) );

%------------------------------------------------------------------
% Compute odds ratio: prob(seg) / prob(no_seg) 
%------------------------------------------------------------------
odds_21 = sum( exp( log_prob - log_prob_noseg ) );

if ~isfinite( odds_21 )
   odds_21 = 1000000; 
end

\end{verbatim}

\subsection{Appendix B: Make Bayesian Blocks}

\DS

This Appendix includes MatLab code for
the iterative procedure to find a multiple change 
point, as
described in \S  \ref{blocks} of the text.

\SS
\begin{verbatim}

function   [ n_seg_vec, xx_vec ] = make_segments( photon_times )
% function [ n_seg_vec, xx_vec ] = make_segments( photon_times )
%
% Input:  photon_times -- photon arrival times, in microseconds
%
% Output: n_seg_vec  -- array of changepoint times
%         xx_vec     -- count rates (c/usec) in the corresponding segments
%
% Note: t_seg = photon_times( n_seg_vec ) generates the changepoint times
%
% ------------------------------------------------------------------------

global prior_ratio min_photons

n_times = length( photon_times );
  min_time = photon_times(       1 );
  max_time = photon_times( n_times );
    delta_t  = ( max_time - min_time ) / ( n_times - 1 );
    min_tick = floor( min_time - 0.5 * delta_t );
    max_tick =  ceil( max_time + 0.5 * delta_t );
    n_ticks = max_tick - min_tick + 1; % Number of microsecond "ticks"

nseg_1_vec = [ 1 ];
nseg_2_vec = [ n_times ];
nosubs_vec = [ 0 ];
xx_vec = [ n_times / n_ticks ];
no_seg_flag = 0;

while no_seg_flag == 0

   num_segments = length( nseg_1_vec );
   no_seg_flag = 1; % set escape unless do a sub-segmentation

   nseg_1_work = [];
   nseg_2_work = [];
   nosubs_work = [];
   xx_work  = [];

   for seg_id = 1: num_segments

      do_it = 1; % do this one, unless ...

      % ... this one has been done before!
      if nosubs_vec( seg_id ) == 1
         do_it = 0;
      end

      nseg_1 = nseg_1_vec( seg_id );
      nseg_2 = nseg_2_vec( seg_id );
      x_seg  = xx_vec( seg_id );

      times_this = photon_times( nseg_1: nseg_2 );
      nt_this = length( times_this );

      if do_it > 0

         % Determine previous time
         time_this_1 = times_this(1);

         if time_this_1 == photon_times(1);

            % Special handling for first point in full array,
            %   or if it is the second point, but the first two
            %   (or more) times are equal:
            ii = find( times_this > time_this_1 );
            if isempty(ii) 
               delt_t = 2; % Token value
            else
               delt_t = times_this(ii(1)) - time_this_1;
            end
            t_0 = time_this_1 - delt_t;

         else

            % t_0 is the time just previous to the sub-array
            t_0 = photon_times( nseg_1 - 1 );

         end

         % Determine subsequent time
         time_this_n = times_this(nt_this);

         if time_this_n == photon_times(n_times);

            % Special handling for last point in full array,
            %   or if it is the second-to-last point, but the 
            %   last two (or more) times are equal:
            ii = find( times_this < time_this_n );
            if isempty(ii) 
               delt_t = 2; % Token value
            else
               delt_t = time_this_n - times_this(ii(length(ii)));
            end
            t_n = time_this_n + delt_t;

         else

            % t_n is the time just after the sub-array
            t_n = photon_times( nseg_2 + 1 );

         end

         [ n_seg, odds_ratio, log_prob ] = find_change( times_this, t_0, t_n );

         % ... one of the proposed subsegments is too short:
         n_seg_right = nt_this - n_seg;
         if (n_seg <= min_photons) | (n_seg_right <= min_photons)
            do_it = 0; 
         end

         % ... the significance criterion not met:
         if odds_ratio < prior_ratio
            do_it = 0; 
         end

      end

      if do_it > 0

         % Subsegment this one; do not escape yet
         no_seg_flag = 0; 

         ntimes_1_left  = nseg_1;
         ntimes_1_right = nseg_1 + n_seg - 1;

         ntimes_2_left  = nseg_1 + n_seg;
         ntimes_2_right = nseg_2;

   
         n_ticks_left  = times_this( n_seg   ) - times_this(     1 ) + 1;
         n_ticks_right = times_this( nt_this ) - times_this( n_seg ) + 1;

         nn_left  = n_seg;
         nn_right = nt_this - n_seg;

         x_seg_left  = nn_left  / n_ticks_left;
         x_seg_right = nn_right / n_ticks_right;

         nseg_1_work = [ nseg_1_work ntimes_1_left ntimes_1_right ];
         nseg_2_work = [ nseg_2_work ntimes_2_left ntimes_2_right ]; 
         xx_work     = [ xx_work  x_seg_left  x_seg_right  ]; 
         nosubs_work = [ nosubs_work 0 0 ];

      else

         % No sub-segmenting of this segment;
         % so just stuff in the beginning, end, mark
         % as "nosubs" so that it will not be done again
         nseg_1_work = [ nseg_1_work nseg_1 ];
         nseg_2_work = [ nseg_2_work nseg_2 ]; 
         xx_work  = [ xx_work x_seg ]; 
         nosubs_work = [ nosubs_work 1 ];

      end

   end

   % Post the segmentations just done:
   nseg_1_vec = nseg_1_work;
   nseg_2_vec = nseg_2_work;
   xx_vec  = xx_work;
   nosubs_vec = nosubs_work;

end
 
n_seg_vec = nseg_2_vec;


\end{verbatim}
\DS

\eject


\begin{thebibliography}{}

\bibitem[Abry and Flandrin (1996)]{abry_2}
Abry, P., \& Flandrin, P. 1996,
``Point Processes, Long-Range Dependence and Wavelets,''
in Wavelets in Medicine and Biology, 
eds. A. Aldroubi \& M. Unser, CRC Press,
Boca Raton, 413


\bibitem[Abry, Goncalves and Flandrin (1995)]{abry_1}
Abry, P., Goncalves, P., \& Flandrin, P. 1995,
``Wavelets, spectrum analysis and 1/f processes,''
in Lecture Notes in Statistics, No. 103, Wavelets and Statistics,
eds. A. Antoniadis \& G. Oppenheim, Springer-Verlag, 15


\bibitem[ Bernaola-Galv\'{a}n, Rom\'{a}n-Rold\'{a}n, and Oliver (1996)]{bernaola}
Bernaola-Galv\'{a}n, Pedro, Rom\'{a}n-Rold\'{a}n, Ram\'{o}n, 
\& Oliver, Jos\'{e} L. 1996,
%``Compositional segmentation and long-range fractal correlations in DNA sequences,''
Phys. Rev. E, 53, 5181


\bibitem[Billingsley (1986)]{billingsley}
Billingsley, P., 1986, Probability and Measure,
John Wiley \& Sons, New York



\bibitem[Brillinger (1977)]{brill_1}
Brillinger, David R., 1997, 
``Some Wavelet Analyses of Point Process Data,''
Proc. 31st Asilomar Conference
on Signals, Systems and Computers.
\verb+http://stat-www.berkeley.edu/users/brill/papers.html+


\bibitem[Brillinger (1978)]{brill_2}
Brillinger, David R., 1978, 
``Comparative Aspects of the Study of Ordinary 
Time Series and of Point Processes,''
pp. 33-133,
Developments in Statistics, Vol. 1, Academic Press, Inc.


\bibitem[Buckheit and Donoho (1995)]{buck_donoho}
Buckheit, J., \& Donoho, D., 1995,
``WaveLab and Reproducible Research,''
in  Wavelets and Statistics,
eds. Antoniadis, A. \& Oppenheim, G., 
Springer Lecture Notes in Statistics, No. 103;
see also
\verb+http://stat.Stanford.EDU/reports/donoho/+
and
\verb+http://sepwww.stanford.edu/research/redoc/+


\bibitem[Chib (1996)]{chib}
Chib, S., 1996, 
``Estimation and Comparison of Multiple Change Point Models,''
Biometrika, in press


\bibitem[Coifman and Donoho (1995)]{coifman_donoho}
Coifman, R. R., \& Donoho, D. L. 1995,
``Translation-Invariant De-Noising,''
in Wavelets and Statistics,
Anestis Antoniadia \& Georges Oppenheim, eds., 
Springer-Verlag Lecture Notes


\bibitem[Donoho (1994)]{donoho_1}
Donoho, David L. 1994, 
``On Minimum Entropy Segmentation,''
Wavelets: Theory, Algorithms, and Applications,
  ed. Chui, C.K., Montefusco, L., \& Pucciio, L.,
Academic Press: New York, 233


\bibitem[Gregory and Loredo (1986)]{gregory_loredo}
Gregory, P. C., \& Loredo, T. J. 1992,
\apj, 398, 146


\bibitem[Gustafsson (1998a)]{gustafsson_1}
Gustafsson, F. 1998a,
``Segmentation of signals using piecewise constant
linear regression,''
submitted to IEEE Transactions on Signal Processing.


\bibitem[Gustafsson (1998b)]{gustafsson_2}
Gustafsson, F. 1998b,
``A Change Detection and Segmentation Toolbox for Matlab,''
Link\"{o}ping University Technical Report LiTH-ISY-R-1669
\verb+http://www.control.isy.liu.se/cgi-bin/reports?author~Gustafsson+

\bibitem[Haight (1967)]{haight}
Haight, F. A., 1967, Handbook of the Poisson Distribution,
John Wiley \& Sons, New York


\bibitem[Jaynes (1997)]{jaynes}
Jaynes, E. T., 1997, 
Probability Theory: The Logic of Science, 
available electronically at \verb+http://bayes.wustl.edu+


\bibitem[Jeffrey (1995)]{jeffrey}
Jeffrey, A. 1995,
Handbook of Mathematical Formulas and Integrals,
Academic Press: New York.


\bibitem[Kolaczyk (1997)]{eric} 
Kolaczyk, Eric D., 1997, 
%``Nonparametric Estimation of Gamma-Ray Burst Intensities Using Haar Wavelets,''
\apj, 483, 34

\bibitem[Lee (1997)]{lee} 
Lee, P., 1997, 
Bayesian Statistics: An Introduction,
John Wiley \& Sons


\bibitem[Loredo (1992)]{loredo} 
Loredo, T.J., 1992,
``Promise of Bayesian Inference for Astrophysics,''
in Statistical Challenges in Modern Astronomy, ed. Feigelson 
and Babu, Springer-Verlag: New York, 275


\bibitem[Marsden and Rothschild (1997)]{marsden_rothschild} 
Marsden, D., \& Tothschild, R.E. 1997,
``Detection of Bursts in Time Series Data Using Bayesian Techniques,''
preprint


\bibitem[Norris et al. (1996)]{attributes} 
Norris, J. P.,
Nemiroff, R. J., Bonnell, J. T., Scargle, J. D.,
Kouveliotou, C., Paciesas, W. S., Meegan, C. A., and
Fishman, G. J. 1996, 
%``Attributes of Pulses in Long Bright Gamma-Ray Bursts,''
\apj, 459, 393


\bibitem[Ogden (1997)]{ogden_1} 
Ogden, R. Todd, 1997,
``Wavelets in Bayesian Change-Point Analysis,''
preprint


\bibitem[Ogden and Parzen (1997a)]{ogden_parzen_a} 
Ogden, R. Todd \& Parzen, Emanuel 1997,
``Data Dependent Wavelet Thresholding in
Nonparametric Regression with Change-point Applications,''
preprint


\bibitem[Ogden and Parzen (1997b)]{ogden_parzen_b} 
Ogden, R. Todd \& Parzen, Emanuel 1997,
``Change-point Approach to Data Analytic 
Wavelet Thresholding,''
preprint


\bibitem[O'Hagan (1994)]{o_hagan} 
O'Hagan, A., 1994, 
Kendall's Advanced Theory of
Statistics: Bayesian Inference, Volume 2B,
John Wiley \& Sons


\bibitem[\`{O} Ruanaidh and Fitzgerald (1996)]{ruanaidh} 
\`{O} Ruanaidh, J. J. \& Fitzgerald, W. J., 1996,
Numerical Bayesian Methods Applied 
to Signal Processing,
Springer: New York



\bibitem[Phillips and Smith (1996)]{phillips_smith} 
Phillips, D. B., \& Smith, A, 1996, 
``Bayesian Model Comparison via Jump Diffusions''
in Markov Chain Monte
Carlo in Practice, ed.\ W.~R.~Gilks, S.~Richardson, \& D.~J.~Spiegelhalter,
Chapman \& Hall, London (1996), 215


\bibitem[Raftery and Akman (1986)]{raftery_akman}
Raftery, A. E., \& Akman, V.E. 1986,
%``Bayesian analysis of a Poisson process with  a change-point,''
Biometrika, 73, 85


\bibitem[Scargle (1997)]{wavelet_methods} 
Scargle, J., 1997, 
``Wavelet Methods in Astronomical Time
Series Analysis,''
in Applications of Time Series Analysis in Astronomy and Metrology,
Chapman \& Hall, 226


\bibitem[Scargle, Norris and Bonnell (1997)]{attributes_2} 
Scargle, J. D., Norris, J, \& Bonnell, J. T., 1997, 
``Attributes of Gamma-Ray Burst Pulses:
I. Short Bursts Analyzed with BATSE Time-Tagged Event Data,'' 
in preparation


\bibitem[Scargle, Steiman-Cameron, Young, Donoho, Crutchfield and Imamura (1993)]{scox} 
Scargle, J., Steiman-Cameron, T., Young, K., Donoho, D.,
Crutchfield, J., \& Imamura, J. 1993, 
%``The Quasi-Periodic Oscillations and Very Low-Frequency
%Noise of Scorpius X-1 as Transient Chaos: A Dripping Handrail?''
ApJ, 411, L91


\bibitem[Schuster (1988)]{schuster} 
Schuster, H. 1988, 
Deterministic Chaos (VCH, New York)


\bibitem[Sivia (1996)]{sivia} 
Sivia, D.S. 1996,
Data Analysis: A Bayesian Tutorial,
Clarendon Press: Oxford


\bibitem[Smith (1975)]{smith} 
Smith, A. F. M., 1975,
%``A Bayesian approach to inference about a change-point
%in a sequence of random variables,''
Biometrika, 62, 407

\bibitem[Stark, Fitzgerald and Hladky (1997)]{stark} 
Stark, J., Fitzgerald, W., \& Hladky, S. 1997,
``Multiple-order Markov Chain Monte Carlo Sampling Methods
with Application to a Changepoint Model,''
Technical Report CUED/F-INFENG/TR. 302,
http://www2.eng.cam.ac.uk/~jas/pubs.html

\bibitem[Stoyan, Kendall, and Mecke (1995)]{stoyan} 
Stoyan, D., Kendall, W., and Mecke, J. (1995),
Stochastic Geometry and its Applications, 
2nd edition, John Wiley \& Sons: New York

\bibitem[Sugiura and Ogden (1997)]{sugiura_ogden} 
Sugiura, N. \& Ogden, R. T., 1997,
``Testing Change-points with Linear Trend,''
preprint


\bibitem[West and Ogden (1997)]{west_ogden} 
West, R. Webster, \& Ogden, R. Todd, 1997,
``Continuous-time Estimation of a Change-point in a Poisson Process,''
preprint


\bibitem[Worsley (1986)]{worsley} 
Worsley, K. J. 1986, 
``Confidence regions and tests for a change-point in a 
sequence of exponential family random variables,''
Biometrika, 73, 91


\bibitem[Young and Scargle (1996)]{young_scargle} 
Young, K., \& Scargle, J. D. 1996,
\apj, 468, 617


\end{thebibliography}
\end{document}